\documentclass[prd,twocolumn,preprintnumbers,superscriptaddress,nofootinbib]{revtex4}
\pdfoutput=1
\usepackage{textcomp} 
\usepackage{slashed}
\usepackage{epsfig,latexsym,cancel,amssymb,amsmath,verbatim,mathrsfs}
\usepackage{color}
\usepackage{graphicx}
\usepackage{enumitem}
\usepackage{graphicx}
\usepackage{amsmath}
\usepackage{amssymb,bm,mathrsfs,bbm,amscd}
\usepackage{color}
\usepackage{subfigure}
\usepackage{lipsum}
\usepackage[fulladjust]{marginnote}
\usepackage[final]{pdfpages}
\usepackage[hyperfootnotes=false]{hyperref}
\usepackage[normalem]{ulem}
\usepackage{float}
\usepackage{placeins}
\usepackage{xspace}
\usepackage{units}
\usepackage{gensymb}
\usepackage{cancel}

\newcommand{\beq}{\begin{equation}}
\newcommand{\eeq}{\end{equation}}
\newcommand{\be}{\begin{equation}}
\newcommand{\ee}{\end{equation}}
\newcommand{\bea}{\begin{eqnarray}}
\newcommand{\eea}{\end{eqnarray}}

\newcommand{\eqal}[1]{\begin{align}#1\end{align}}

\allowdisplaybreaks[4]

\begin{document}

\title{Investigating Higgs self-interaction through di-Higgs plus jet production}

\author{Kangyu Chai}
\email{chaikangyu@ihep.ac.cn}
\affiliation{Theoretical Physics Division, Institute of High Energy Physics, Chinese Academy of Sciences, Beijing 100049, China}
\affiliation{School of Physics, University of Chinese Academy of Science, Beijing 100049, China}

\author{Jiang-Hao Yu}
\email{jhyu@itp.ac.cn}
\affiliation{CAS Key Laboratory of Theoretical Physics, Institute of Theoretical Physics, Chinese Academy of Sciences, Beijing 100190, China}
\affiliation{School of Physical Sciences, University of Chinese Academy of Sciences, Beijing 100049, China}
\affiliation{Center for High Energy Physics, Peking University, Beijing 100871, China}
\affiliation{School of Fundamental Physics and Mathematical Sciences, Hangzhou Institute for Advanced Study, UCAS, Hangzhou 310024, China}
\affiliation{International Centre for Theoretical Physics Asia-Pacific, Beijing/Hangzhou, China}

\author{Hao Zhang}
\email{zhanghao@ihep.ac.cn}
\affiliation{Theoretical Physics Division, Institute of High Energy Physics, Chinese Academy of Sciences, Beijing 100049, China}
\affiliation{School of Physics, University of Chinese Academy of Science, Beijing 100049, China}
\affiliation{Center for High Energy Physics, Peking University, Beijing 100871, China}

\date{\today}

\begin{abstract}
The Higgs self coupling measurement is quite essential for determining the shape of the Higgs potential and nature of the Higgs boson. We propose the di-Higgs plus jet final states at hadron colliders to increase the discovery sensitivity of the Higgs self coupling at the low invariant mass region. Our simulation indicates that the allowed region of the Higgs self coupling would be further narrowed from $[-1.5,6.7]$ from the most recent ATLAS report down to $[0.5, 1.7]$. Furthermore, we find negative Higgs self couplings would be disfavored beyond $2\sigma$ confidence level at a future 100\,TeV collider with the help of this signal. 
\end{abstract}
\maketitle

\section{Introduction}
The discovery of the Higgs boson in 2012\,\cite{ATLAS:2012yve,CMS:2012qbp} represents one milestone of modern particle physics. It provides the evidence that the observed Higgs boson is the one predicted by the Standard Model (SM). While the SM parameters have essentially been measured to a very high precision level, the Higgs self couplings, important for electroweak symmetry breaking and understanding its connection to other fundamental questions like electroweak baryogenesis\,\cite{Morrissey:2012db}, have not been measured directly yet. More importantly, depending on the nature of the Higgs boson, such as fundamental, pseudo-Goldstone, pseudo-Dilaton, or partially composite, the shape of the Higgs potential could be quite different from the SM one\,\cite{Agrawal:2019bpm}. Indeed, a wide range of new physics (NP) models beyond the SM predict modified Higgs potentials that lead to $\mathcal{O}(1)$ corrections to the Higgs self couplings, the Coleman-Weinberg\,\cite{Hill:2014mqa,Helmboldt:2016mpi,Hashino:2015nxa} and the tadpole-induced\,\cite{Galloway:2013dma,Chang:2014ida} Higgs scenarios for example. Therefore, a precision measurement of the Higgs self couplings would provide an important benchmark for model identification and deepen our understanding on electroweak symmetry breaking (EWSB).

Experimentally, the Higgs self couplings could be measured directly from Higgs pair production or Higgs associated production. Due to their lower cross sections for the latter, in this work, we focus specifically on the former that is dominated by gluon-gluon fusion (ggF) at hadron colliders that has been studied in detail earlier\,\cite{ATLAS:2017muo,CMS:2017cwx,Contino:2016spe}.\footnote{Lepton colliders could also measure Higgs self couplings directly, see, for example, Refs.\,\cite{Tian:2016qlk,Abramowicz:2016zbo,Barklow:2017awn,DiVita:2017vrr}. We focus on hadron colliders in this work given the foreseen high-luminosity/energy era of the Large Hadron Collider (LHC) in the near future.} However, due to a strong cancellation near the kinematical threshold, the cross sections for Higgs pair production is highly suppressed -- At a $13\,\rm TeV$ $pp$ collider, the ggF cross section for the Higgs pair production is calculated at next-to-next-to leading order in finite top-quark mass approximation, and the result is $31.02^{+2.2\%}_{-5.0\%}(\rm scale)^{+4\%}_{-18\%}(m_{\rm top})\pm3.0\%(\alpha_s+\rm PDF)\rm\,fb$\,\cite{Shao:2013bz,deFlorian:2015moa,DiMicco:2019ngk,Baglio:2020wgt}. Here, ``scale'' stands for the uncertainty from finite order quantum chromodynamics calculation, ``$m_{\rm top}$'' that from the top-quark mass scheme\,\cite{Grazzini:2018bsd,Baglio:2020wgt}, and ``$\alpha_s+{\rm PDF}$'' that from the strong coupling constant and the parton distribution functions. As a consequence, the Higgs self couplings are only very loosely bounded\,\cite{ATLAS:2021ifb}, let alone their precision determination.

Nevertheless, it is worth pointing out that current experimental searches mainly focus on the high di-Higgs invariant mass region, while it is perhaps universally recognized that the it is the low mass region that is most sensitive to NP. This motivates the study of Higgs self couplings in the low mass region in this work. To increase the significance of the di-Higgs signal in this region, we consider instead Higgs pair production through ggF with an extra hard jet in the final state, i.e., $pp\to hh+jet+X$, with $X$ any other particles in the final state that we are not interested in. Similar to the pure di-Higgs production channel, we consider the $bb\gamma\gamma$ decay channel of the Higgs pair for its cleanness and the unambiguity in reconstructing the two Higgs particles.

The rest of the paper is organized as follows: In section\,\ref{sec:setup}, we set up the framework used in this work, and briefly summarize previous searches in di-Higgs production. We then detail our strategy for $pp\to hh+jet+X$ searches in section\,\ref{sec:ourstra}. Results from detector-level simulation for this channel are then presented in section\,\ref{sec:result}, and we conclude in section\,\ref{sec:con}.

\section{Higgs Nature determination via Higgs self interactions}
\label{sec:setup}
In the effective field theory (EFT) framework, new physics effect in the Higgs sector could be described using Higgs EFT (HEFT) and standard model EFT (SMEFT) in the broken and unbroken phase of electroweak symmetry, respectively. Although SMEFT is the most popular EFT scenario, its validity relies on the assumptions that new physics should decouple at low energy scale. On the other hand, the HEFT would describe the Higgs potential in the broken phase and thus describe the nature of the Higgs and the Higgs couplings in a more general way. 

In the HEFT scenario~\cite{Appelquist:1980vg,Longhitano:1980iz,Feruglio:1992wf,Herrero:1993nc,Buchalla:2013rka,Krause:2018cwe,Brivio:2016fzo,Sun:2022ssa}, the electroweak gauge symmetry is broken down to the $U(1)_{\rm em}$ and the global $SU(2)_L \times SU(2)_R/SU(2)_V$ symmetry in the Higgs sector is non-linearly realized. Treating the Higgs boson $h$ as an electroweak singlet, the HEFT Lagrangian at the leading order reads
\eqal{
\mathcal{L} &=  \frac{v^{2}}{4} \operatorname{Tr}\left[D_{\mu} U^{\dagger} D^{\mu} U\right]\left(1+2 a \frac{h}{v}+b \frac{h^{2}}{v^{2}}+\cdots\right) +  \\
&\frac{1}{2}\left(\partial_{\mu} h\right)^{2} -\frac{1}{2} m_{h}^{2} h^{2} - \kappa_\lambda \left(\frac{m_h^2}{2v}\right) h^3 - \kappa_h \left(\frac{m_h^2}{8v^2}\right) h^4 + \cdots \nonumber
}
which parametrize the Higgs potential in the polynomial form and does not depends on the decoupling behavior. Depending on the nature of the Higgs boson, the Higgs potential could be different from the SM form as parameterized by $\kappa_{\lambda,h}$.

In the SMEFT scenario~\cite{Weinberg:1979sa, Buchmuller:1985jz, Grzadkowski:2010es, Lehman:2014jma, Henning:2015alf, Liao:2016hru, Li:2020gnx, Murphy:2020rsh, Li:2020xlh, Liao:2020jmn,Li:2022tec}, the Higgs potential can be expressed as
\eqal{
V_h\supset-\mu^{2} H^{\dagger} H+\lambda\left(H^{\dagger} H\right)^{2}+\frac{c_{6}}{\Lambda^{2}} \lambda\left(H^{\dagger} H\right)^{3}+\cdots
}
where $\Lambda$ is the UV cutoff, $c_6$ is some dimensionless Wilson coefficient, and ``$\cdots$'' represents some higher dimensional operators of the SMEFT. The triple and quartic Higgs couplings can then be easily matched to above parameters after electroweak symmetry breaking upon substituting $H$ for $(0,v+h)^{\rm T}/\sqrt{2}$, leading to\,\cite{Agrawal:2019bpm}
\eqal{
V_h\supset &\, \frac{1}{2}\left( 2\lambda  v^2 + \frac{3 c_6 \lambda  v^4}{\Lambda ^2} \right)h^2 + \lambda v  \left(1 + \frac{5 c_6 v^2}{2 \Lambda ^2} \right)  h^3\nonumber\\
&\,+ \frac{1}{4} \lambda  \left( 1 + \frac{15 c_6 v^2}{2\Lambda ^2} \right) h^4 + \cdots\nonumber 
}
where we have applied the minimization condition $\mu^2 = \lambda v^2 + 3c_6\lambda v^4/(4\Lambda^2)$ to obtain the expression above and discarded terms that are not interested for the study in this work. Matching between the HEFT and the SMEFT operators, the Higgs mass and the $\kappa$'s are defined as, up to $\mathcal{O}(1/\Lambda^2)$,
\small{
\eqal{
m_h^2 \equiv 2\lambda  v^2 + \frac{3 c_6 \lambda  v^4}{\Lambda ^2}, \kappa_\lambda \equiv 1 + \frac{c_6 v^2}{\Lambda^2}, \kappa_h \equiv 1 + \frac{6c_6 v^2}{\Lambda^2}.
}
} Note that one reproduces SM tree-level results upon setting $c_6=0$. We comment on that $(H^{\dagger} H)\Box(H^{\dagger} H)$ and $(H^{\dagger} D^\mu H)^*(H^{\dagger} D^\mu H)$ would also contribute to shifting the Higgs mass and the Higgs self couplings from the kinetic Lagrangian, we leave out these operators in our analysis since they are highly constrained by electroweak precision physics and/or $hVV\,(V=W^\pm,Z)$ couplings\,\cite{Agrawal:2019bpm}.

\begin{table}\caption{{Higgs self couplings $\kappa_\lambda$ and $\kappa_h$ in different cases. Here, ``$\rm MCH_{5+5}$'' means the minimal composite Higgs model\,\cite{Agashe:2004rs,Contino:2006qr}, ``$\rm CTH_{8+1}$'' the composite twin Higgs model\,\cite{Geller:2014kta,Barbieri:2015lqa,Low:2015nqa}, and ``CW'' the Coleman-Weinberg Higgs scenario\,\cite{Hill:2014mqa,Helmboldt:2016mpi,Hashino:2015nxa}. The first (second) subscript of the model name represents the fundamental representation of the left-(right-)handed top quark under the global symmetry, which is SO(5) and SO(8) for ``$\rm MCH_{5+5}$'' and ``$\rm CTH_{8+1}$'', respectively. In the CW Higgs scenario, numbers in parentheses are results up to the two-loop order from Refs.\cite{Hill:2014mqa,Helmboldt:2016mpi}.}}\label{tab:coupling}
 \begin{center}
  \begin{tabular}{c|c|c}
   \hline\hline
  Higgs self couplings & $\kappa_\lambda$ & $\kappa_h$ \\
   \hline  
  SM  & $1$ & $1$ \\ 
  \hline
  SMEFT (with $\mathcal{O}_6$) & $1 + \frac{c_6v^2}{\Lambda^2}$ & $1 + \frac{6 c_6 v^2}{\Lambda^2}$ \\  
   \hline
  $\text{MCH}_{5+5}$ & $ 1-\frac{3}{2}\xi$ & $ 1-\frac{25}{3}\xi$ \\
   \hline
   $\text{CTH}_{8+1}$ & $ 1-\frac{3}{2}\xi$ & $ 1-\frac{25}{3}\xi$ \\
   \hline
   CW Higgs (doublet) & $\frac{5}{3}(1.75)$ & $\frac{11}{3}(4.43)$ \\
   \hline
   CW Higgs (singlets) & $\frac{5}{3}(1.91)$ & $\frac{11}{3}(4.10)$ \\
   \hline
   Tadpole-induced Higgs & $\simeq 0$ & $\simeq 0$ \\
   \hline\hline
  \end{tabular}
 \end{center}
\end{table}

Depending on the nature of the Higgs boson, the Higgs boson could be fundamental, pseudo-Goldstone, pseudo-Dilaton, or partially composite due to strong dynamics condensation\,\cite{Agrawal:2019bpm}. For {a} fundamental Higgs {boson}, such as the SM {Higgs} {boson} and {its} scalar/gauge extensions, and supersymmetric models, the form of the Higgs potential {is} polynomial on the Higgs doublet. {In this case,} there {usually exist} additional scalars mixed with the {SM} Higgs boson, thus modifying the SM Higgs self couplings with {some} enhancement. {In contrast,} if the Higgs boson is pseudo-Goldstone due to the vacuum misalignment, the curvature of the Higgs field would cause the Higgs couplings {to be} always smaller than {their} SM values. On the other hand, if the Higgs boson is a pseudo-dilaton, the Higgs potential would be {of} purely {the} Coleman-Weinberg type and thus the Higgs self-couplings would be larger than the SM ones. Finally, if the symmetry breaking is partially induced by condensation, it is possible to have the tadpole-induced symmetry breaking and thus the Higgs self couplings are nearly zero. We summarize the Higgs self couplings in different scenarios {discussed above in} Table\,\ref{tab:coupling}.  

\begin{figure*}[!hbtp]
\caption{\label{fig:FD} Leading order Feynman diagrams for the $gg\rightarrow hhg$ process.}
\centering
\includegraphics[width=0.9\linewidth]{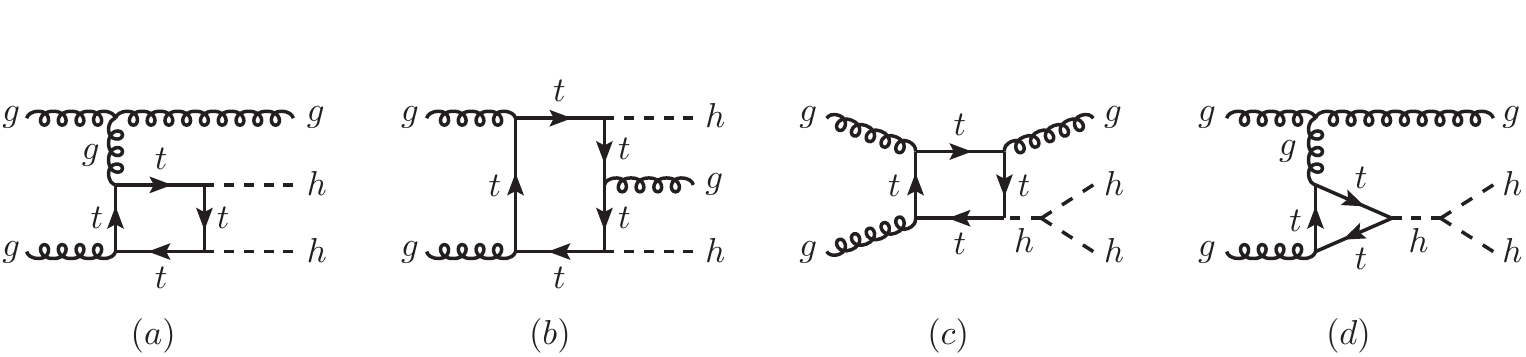}
\end{figure*}

Therefore, measuring the Higgs self couplings could possibly unveil the pattern of EWSB, which in turn helps determine the nature of the Higgs boson. In this context, Higgs boson pair production $pp\to hh+X$ through ggF plays a key role due to its direct sensitivity to $\kappa_\lambda$ and relatively large production cross section.\footnote{Other production channels such as vector-boson fusion, $t\bar{t}/W/Z$ or single-top associated production, also offer the opportunity for measuring Higgs self couplings directly. We choose not to adopt these channels for the discussion in this work due to their lower cross sections\,\cite{Frederix:2014hta}.} Various final states of $hh$ have been considered previously, with the promising ones including $b\bar{b}\gamma\gamma$\,\cite{Baur:2003gp,Barger:2014qva,Alves:2017ued,Adhikary:2017jtu,Kim:2018uty}, $b\bar{b}\tau^\pm\tau^\mp$\,\cite{Baur:2003gpa,Dolan:2012rv,Barr:2013tda}, $b\bar{b}W^\pm W^\mp$\,\cite{Papaefstathiou:2012qe}, $b\bar{b}b\bar{b}$\,\cite{FerreiradeLima:2014qkf,Wardrope:2014kya,Behr:2015oqq}, and $W^\pm W^\mp W^\pm W^\mp$\,\cite{Baur:2002rb,Baur:2002qd,Li:2015yia}. Among them, $b\bar{b}\gamma\gamma$ has been recognized as the most promising channel for precision Higgs boson self coupling measurement thanks to its clean final states and unambiguity in reconstructing the Higgs bosons with the decay products of $hh$. Experimentally, this channel has been intensively investigated at the LHC\,\cite{ATLAS:2018dpp,CMS:2018tla,ATLAS:2019qdc,CMS:2020tkr}, and recently, the ATLAS collaboration reported their improved results with $-1.5\le\kappa_\lambda\le6.7$ at 95\% confidence level (CL) by considering the full Run 2 data set of 139\,$\rm fb^{-1}$ at 13\,TeV and utilizing the $b\bar{b}\gamma\gamma$ channel\,\cite{ATLAS:2021ifb}. We refer the readers to \cite{ATLAS:2021ifb} for the details of their analysis and outline their strategy below for reference. The preselection cuts they apply are: 
\begin{itemize}
\item $p_{T,\gamma}^{\rm leading}\ge35$\,GeV, $p_{T,\gamma}^{\rm sub-leading}\ge25$\,GeV;
\item At least two photons;\footnote{These photons shall correspond to those reconstructed from topologically connected clusters of energy deposits in the electromagnetic calorimeter with pseudorapidity $|\eta|<2.37$. Those with $1.37<|\eta|<2.37$ in the transition region between the barrel and endcap electromagnetic calorimeters are rejected. Furthermore, to avoid photon misidentification, the calorimeter-based (track-based) isolation needs to be less than 6.5\% (5\%) of the photon transverse energy\,\cite{ATLAS:2021ifb}.}
\item $105<m_{\gamma\gamma}<160$\,GeV;
\item $p_{T,\gamma}^{\rm leading}>0.35m_{\gamma\gamma}$ and $p_{T,\gamma}^{\rm sub-leading}>0.25m_{\gamma\gamma}$;
\item Exactly two $b$-tagged jets;
\item No electrons or muons;
\item Fewer than six jets with $|\eta|<2.5$.
\end{itemize}
Events passed these cuts are then divided into two regions with $m_{b\bar{b}\gamma\gamma}^*<350$\,GeV for and $m_{b\bar{b}\gamma\gamma}^*>350$\,GeV, targeting the SM and the BSM signal, respectively. Here, $m_{b\bar{b}\gamma\gamma}^*$ is defined as $m_{b\bar{b}\gamma\gamma} - m_{b\bar{b}} - m_{\gamma\gamma}+250$\,GeV for the diphoton and $b$-tagged jets system. In each region, the boosted decision tree (BDT) method is adopted for event selection. For the training variables and the event selection criteria in each region, see their Tables 2-4.

While perhaps it is universally acknowledged that the phase space region with small di-Higgs invariant mass $m_{hh}$ is most sensitive to $\kappa_\lambda$, this region is mostly excluded in current experimental analysis, and that motivates the study in this work. To that end, we consider instead Higgs-pair production via ggF with an extra light jet in the final state. The extra hard jet in the final state would boost the transverse momenta of the Higgs pair such that one could gain extra significance to the low $m_{hh}$ region in the end. This in turn helps the determination of the Higgs self couplings as we will see later in this article. We detail our analysis in the next section.

\section{Di-higgs plus jet signature at hadron collider}
\label{sec:ourstra}
As discussed above, we consider $pp\to hh+jet+X$ instead of $pp\to hh+X$ in this work in order to extract the Higgs self couplings from the low $m_{hh}$ region. This relies on the fact that when an additional hard jet is present in the final state, the di-higgs invariant mass would tend to be small due to kinematics. Furhtermore, the additional hard jet would also highly suppress the SM QCD background thanks to its large transverse momentum. All together, the $pp\rightarrow hh+jet+X$ channel could then be a promising candidate to extract $\kappa_{\lambda}$ in small $m_{hh}$ region as we shall see below.

Contributions to $pp\rightarrow hh+jet+X$ mainly arise from the $gg\rightarrow hhg$ channel, whose leading order diagrams in the SM are shown in FIG.\,\ref{fig:FD}. As discussed earlier, we focus on the $hh\rightarrow b\bar{b}\gamma\gamma$ decay channel of the Higgs pair, and study its prospect for $\kappa_\lambda$ extraction at a future $100 \rm TeV$ $pp$ collider due to the limited statistics at the LHC or its high-luminosity era. At parton-level, all the signal and the background events are generated using the five-flavor scheme of \textsf{MadGraph\_aMC@NLO}\cite{MG5}, with the subsequent decay of $h$ done by \textsf{MadSpin}\cite{MadSpin}. The main backgrounds included in this study are
\begin{align}
\nonumber
& pp\rightarrow t\bar{t}\left(h\rightarrow\gamma\gamma\right) \\
\nonumber
& pp\rightarrow t\bar{t}\left(h\rightarrow\gamma\gamma\right)j \\
\nonumber
& pp\rightarrow bb\gamma\gamma j \\
\nonumber
& pp\rightarrow bb\gamma j j \\
\nonumber
& pp\rightarrow bj\gamma\gamma j \\
\nonumber
\end{align}
with $j\in\{g, u, d, s, c, b\}$. All backgrounds are generated using the tree-level event generator of \textsf{MadGraph\_aMC@NLO} to avoid the third background from being the genuine signal. Furthermore, we also apply the following kinematical cuts for event generation:

\begin{figure*}[hbtp]
\caption{\label{fig:sig} Significance distributions for $\kappa_\lambda=0,2,3$ for $pp\rightarrow hh+jet+X$(left panel) and $pp\rightarrow hh+X$(right panel). The significance shows the confidence level (CL) at which one can separate the non-standard scenario with $\kappa_\lambda\neq 1$ from the SM with $\kappa_\lambda=1$.}
\centering
\includegraphics[width=0.49\linewidth]{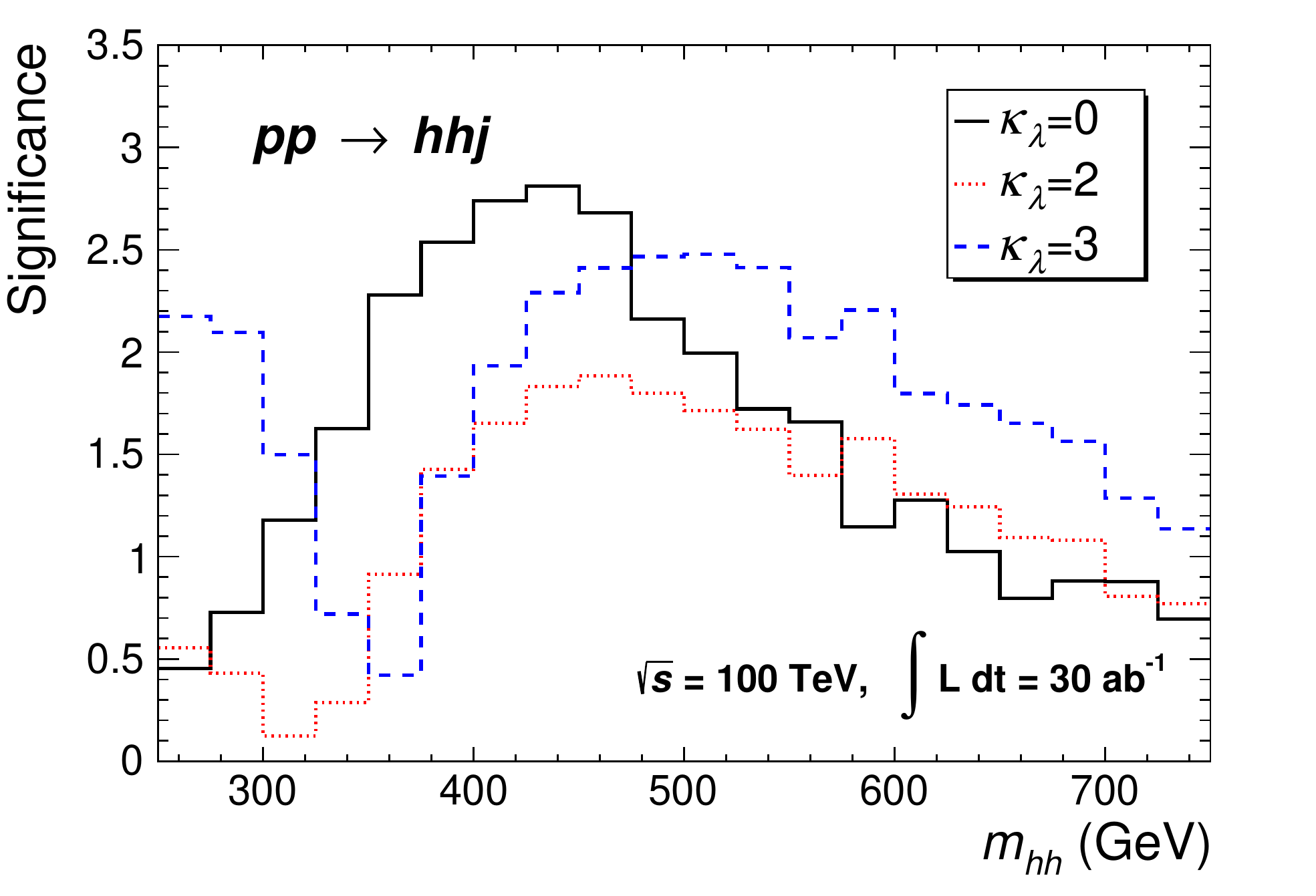}
\includegraphics[width=0.49\linewidth]{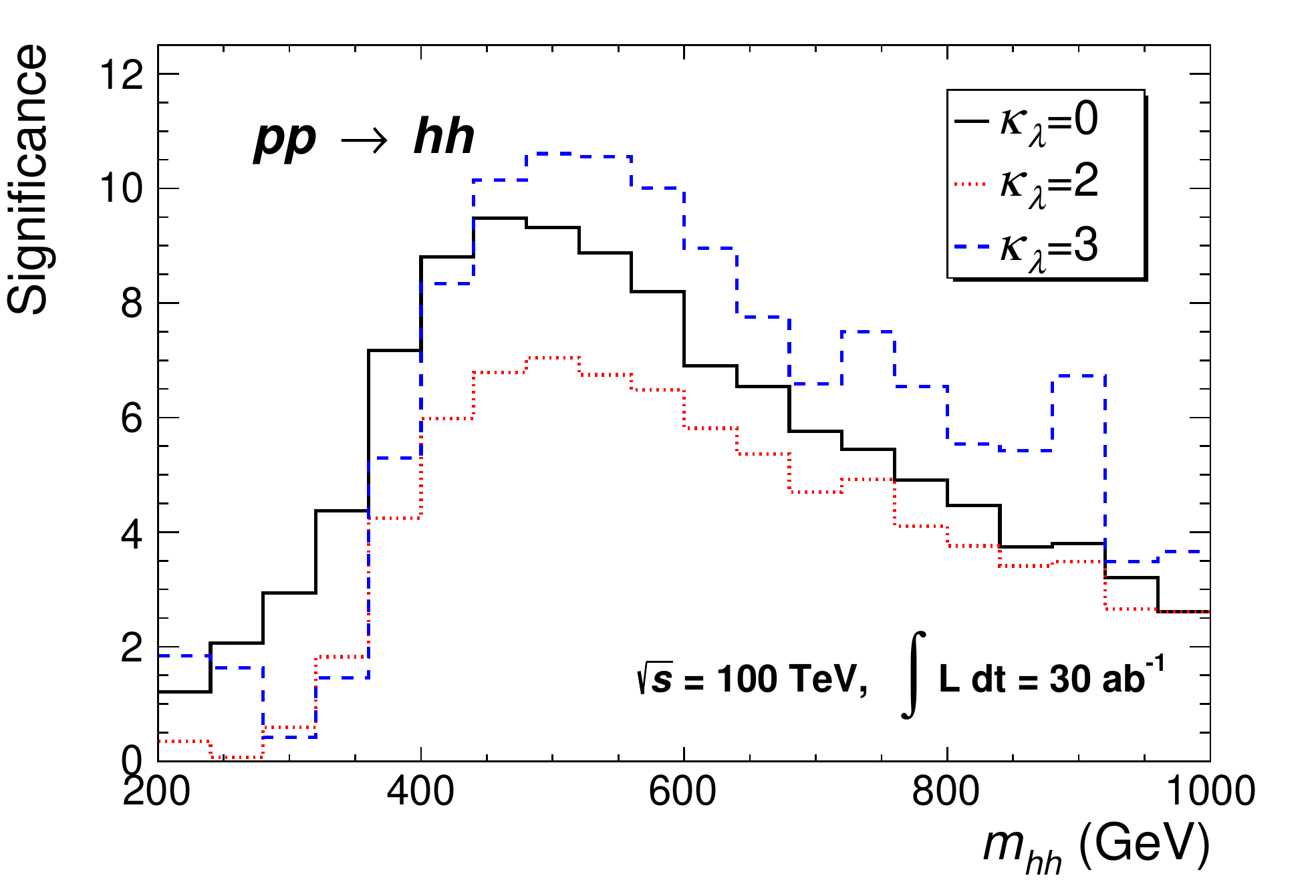}
\end{figure*}

\begin{figure}[!hbtp]
\caption{\label{fig:bkg}Di-Higgs invariant mass distribution for our signal and the SM backgrounds at a future circular $pp$ collider with $\sqrt{s}=100$\,TeV and $\mathcal{L}=30\rm\,ab^{-1}$.}
\includegraphics[width=\linewidth]{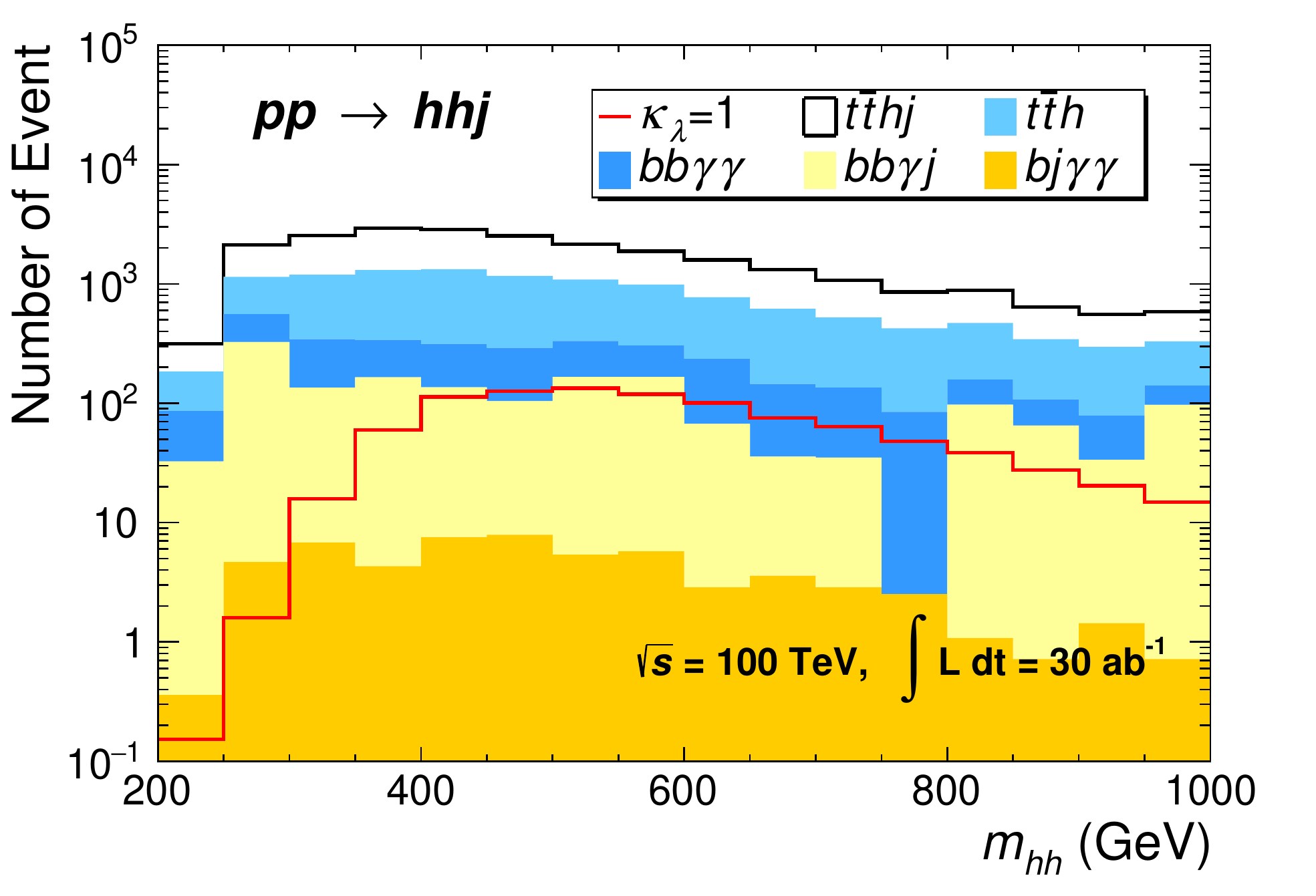}
\centering
\end{figure}
\begin{figure}[hbtp]
\caption{\label{fig:cls}The log-profile-likelihood ratio scanned over $\kappa_\lambda$ for $pp\rightarrow hh+jet+X$ at a future circular 100\,TeV $pp$ collider with $\mathcal{L}=30\rm\,ab^{-1}$.}
\includegraphics[width=\linewidth]{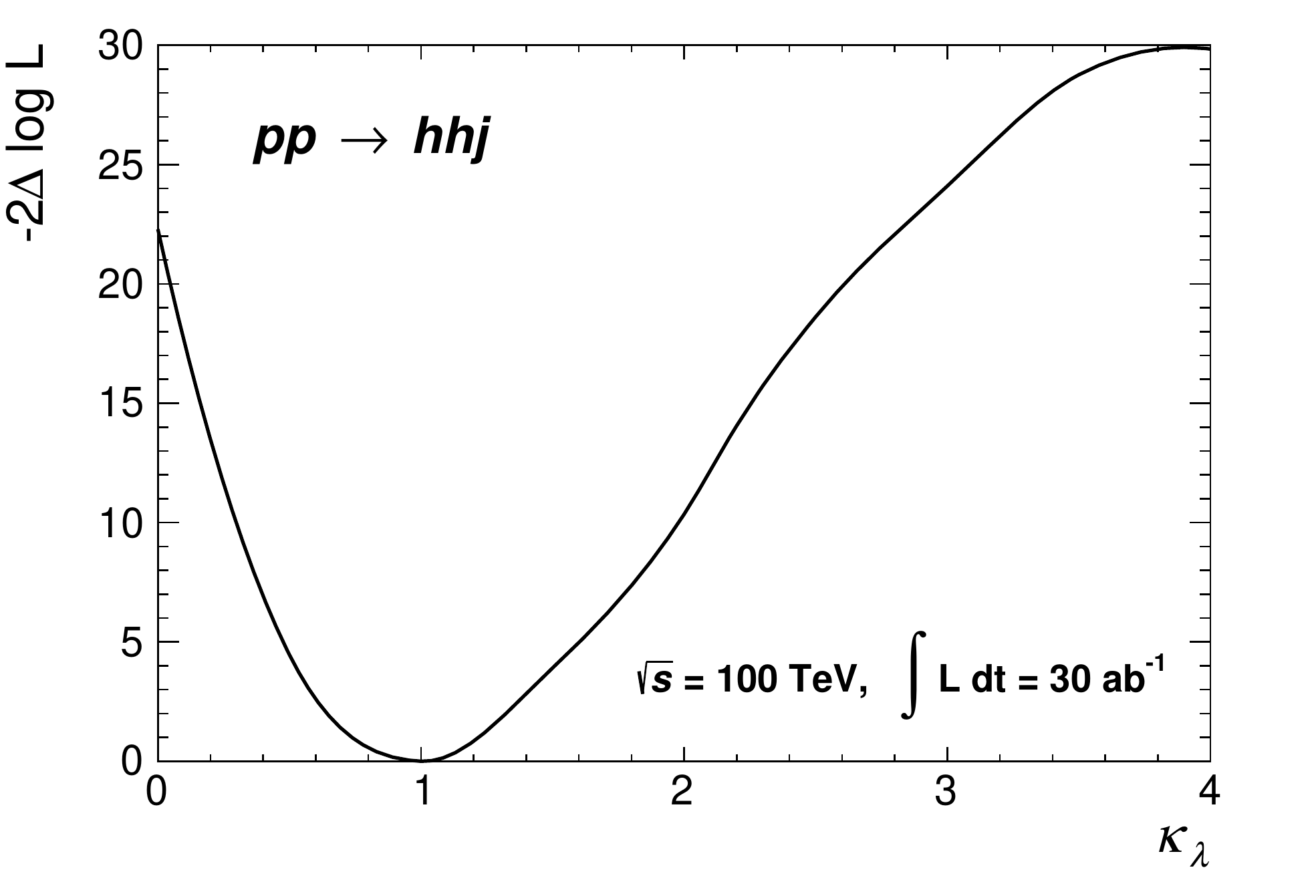}
\centering
\end{figure}

\begin{figure*}[hbtp]
\caption{\label{fig:sigdec} Same as FIG.\,\ref{fig:sig} but obtained by following the analysis in Ref.\,\cite{Goncalves:2018qas} for the right panel and fitting the histograms in FIG.\,\ref{fig:bkg} for the left panel. See the text for details.}
\centering
\includegraphics[width=0.49\linewidth]{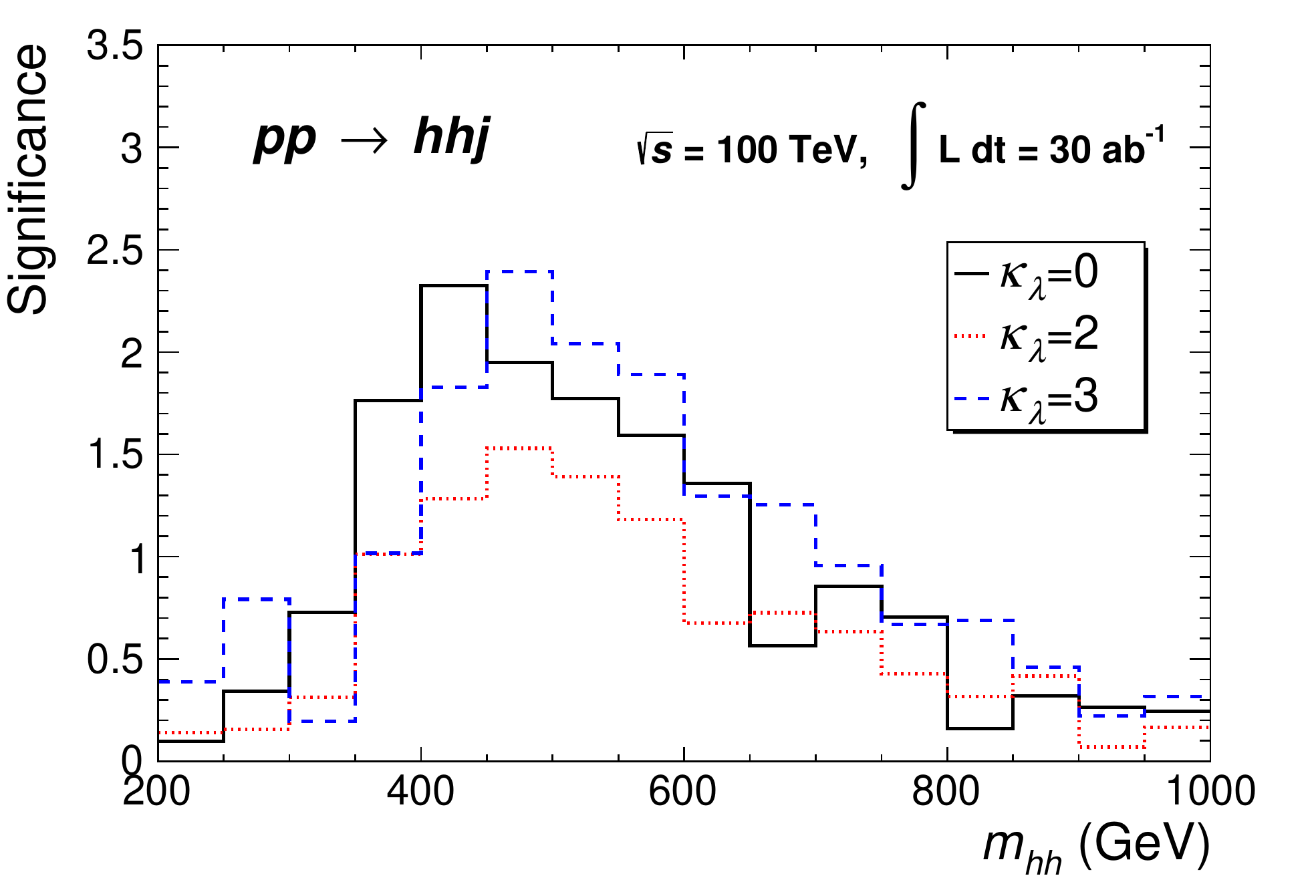}
\includegraphics[width=0.49\linewidth]{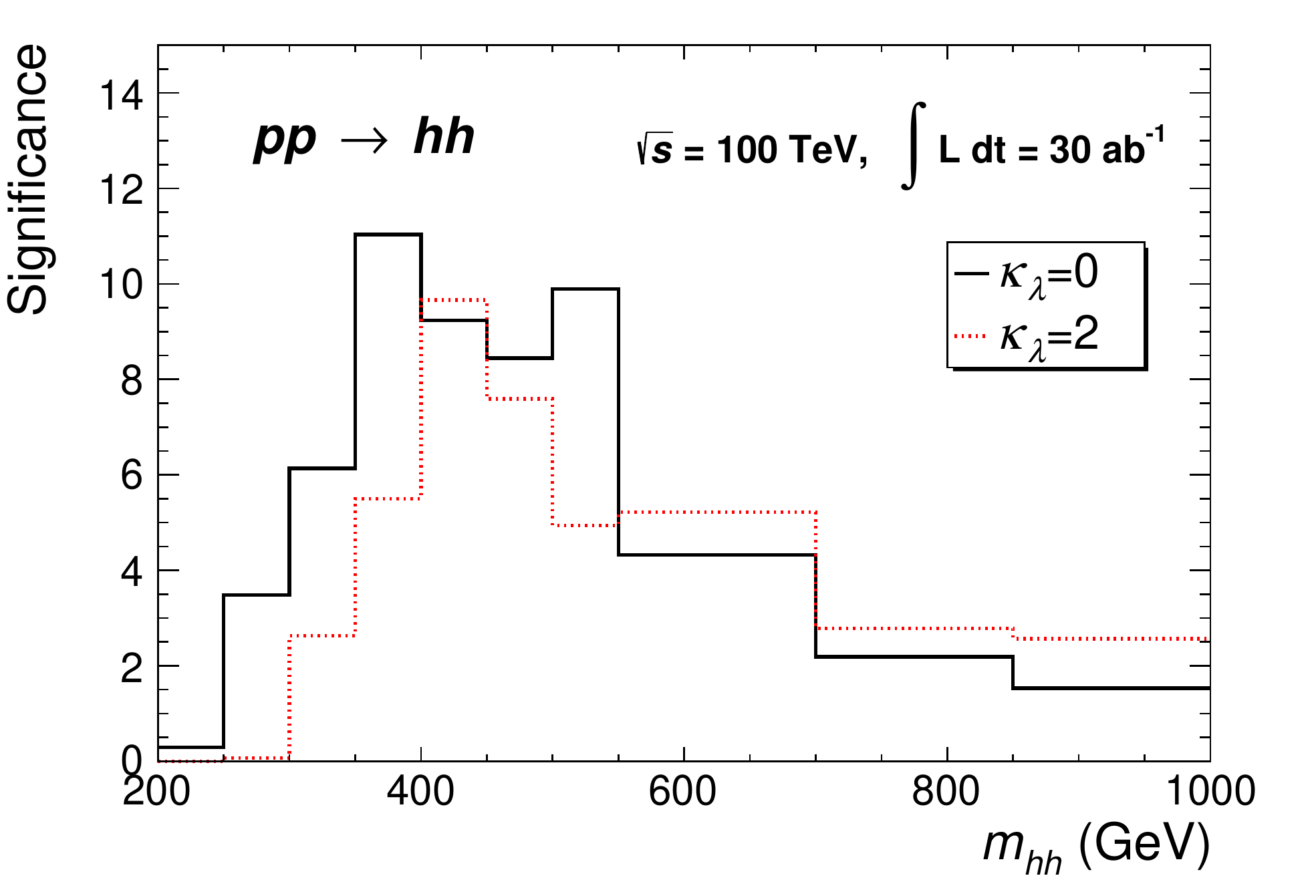}
\end{figure*}

\begin{align}
\nonumber
&\Delta R_{j\gamma,jj,\gamma\gamma} > 0.3 \\\nonumber
&\vert\eta_{b,\gamma}\vert<3, \quad \vert\eta_{i}\vert<5 \\\nonumber
&p_{T,\gamma} > 10 {\rm GeV}, \quad p_{T,j} > 20 {\rm GeV} \\\nonumber
&p^{\rm leading}_{T,j} > {80 \rm GeV} \\\nonumber
& 75{\rm GeV}<m_{bb} < 175{\rm GeV}\\\nonumber
&{100\rm GeV} < m_{\gamma\gamma} < {150\rm GeV}
\nonumber
\end{align}
where $i\in\{g, u, d, s, c\}$. We comment on that cuts on $\Delta R$, $\eta$ and $p_T$ are imposed to avoid infrared divergence. The cuts for $b$-jets and light-flavor jets are applied differently from the fact that the sensitivity region of the detector for $b$-tagging is mostly restricted to $\vert\eta\vert<2.5$. The three exclusive cuts, leading-jet transverse momentum $p_{T,j}^{\rm leading}$, $m_{bb}$ and $m_{\gamma\gamma}$ precisely, are imposed to make our simulation more efficient but still inclusive enough. Additionally, no cuts are put on the decay products of the heavy resonances since otherwise one may underestimate the backgrounds.

For parton-level analysis, the misidentification rate and the smearing effect indicated in\,\cite{Goncalves:2018qas} are employed. For signal event selection, we require exactly two $b$-jets and two photons as in Ref.\,\cite{ATLAS:2021ifb} but with an extra requirement that there be at least one additional jet in the final state. After these preselection cuts, we further apply the following kinematical cuts:
\begin{align}
\nonumber
&\Delta R_{bb,\gamma\gamma,b\gamma}<0.4\\\nonumber
&p_{T,b} > 30 {\rm GeV} ~~ p_{T,\gamma}>30 {\rm GeV}\\\nonumber
&\vert\eta_b\vert<2.5 ~~ \vert\eta_{\gamma}\vert<2.5\\\nonumber
&120 {\rm GeV}<m_{\gamma\gamma}<130{\rm GeV}\\\nonumber
&80 {\rm GeV}<m_{bb}<160 {\rm GeV}\\
& p_{T,j}^{\rm leading}>150 {\rm GeV}
\end{align}
Note that our cuts on $p_{T,\gamma}$ is consistent with those in Ref.\,\cite{ATLAS:2021ifb}, and our range for $m_{\gamma\gamma}$ lies within that of Ref.\,\cite{ATLAS:2021ifb}. After vetoing events not passing above cuts, we display the sensitivity of our signal in the left panel of FIG.\,\ref{fig:sig} as a function of $m_{hh}$ for three benchmarks with $\kappa_\lambda=0,2,3$ in red, blue, and green, respectively. A similar analysis is carried out for the $pp\rightarrow hh+X$ channel based on Ref.\,\cite{Goncalves:2018qas}, and the corresponding results can be seen in the right panel of FIG.\,\ref{fig:sig}. 

In order to show the sensitivity of each channel to different $m_{hh}$ regions, the results are displayed as significance distribution. This distribution is obtained by calculating likelihood ratio $\sqrt{-2\log{(\Lambda/\Lambda_0)}}$ for each bin. 

From the significance distributions at the parton level as shown in FIG.\,\ref{fig:sig}, it is obvious that with an extra hard jet in the final state, the $pp\rightarrow hh+jet+X$ process becomes more sensitive to the Higgs self coupling $\kappa_\lambda$ in the low $m_{hh}$ region. In the meantime, we comment on that the $pp\rightarrow hh+X$ process exhibits a larger significance due to larger statistics, and our signal is relatively more kinematically suppressed due to the hard jet. However, we expect the significance of our signal to be improved, for example, with the BDT method.

\section{Detector-level Simulations}
\label{sec:result}
We now move to the discussion on the detector side. All the parton-level events generated in the previous section are showered by \textsf{Pythia8}\cite{Pythia8} for hadronization, and the detector effect is then simulated using \textsf{Delphes}\cite{Delphes}. Since the full NLO QCD corrections to the $pp\rightarrow hh+jet+X$ process are still missing, no additional $K$-factor will be included in our simulation.

Furthermore, for detector level simulations, the photon efficiency is tuned to be $90\%$ and all jets are reconstructed with the anti-$k_T$ algorithm with jet radius $R=0.4$. The $b$-tagging efficiency is set to be $80\%$, and the mis-tagging rate is set to be $10\%$ for charm-jet and $1\%$ for other light-flavor jets. Also, the jet-faking-photon rate is set to be $0.05\%$. In addition, as a trigger requirement, all photons and $b$-jets should have $p_T>30\rm\,GeV$ and $0<\vert\eta\vert<2.5$, and photons between the barrel and endcap calorimeter, or equivalently, photons with $1.37<\vert\eta_\gamma\vert<1.52$, are excluded for object selection. Then, the $b\bar{b}\gamma\gamma+jet$ final state is reconstructed with exactly two $b$-tagged jets, two photons and at least one additional jet satisfying:
\begin{align}
\nonumber
& 122 {\rm GeV}< m_{\gamma\gamma} < 128 {\rm GeV}, \\
\nonumber
& 95 {\rm GeV} < m_{bb} < 155 {\rm GeV}, \\
\nonumber
& p_{T,j}^{\rm leading} > 150 {\rm\,GeV}, \quad \vert\eta_j\vert<4.5
\end{align}
At this stage, the SM QCD backgrounds are all well suppressed except $t\bar{t}h$ and $t\bar{t}h+jet$. In order to suppress these two backgrounds, any event which contains one or more isolated lepton ($e^\pm,\mu^\pm$) with $p_T > 25 \rm\, GeV$ and $\vert\eta\vert < 2.5$ will be vetoed. Moreover, for events with at least four additional jets, the following quantity is calculated to veto the top quark:
\begin{align}
\label{eq:topness}
\chi^2&=\min\Big\{\frac{\left(m_W-m_{i_1i_2}\right)^2}{\sigma_W^2}+\frac{\left(m_t-m_{i_1i_2j_1}\right)^2}{\sigma_t^2}\\\nonumber
&+\frac{\left(m_W-m_{i_3i_4}\right)^2}{\sigma_W^2}+\frac{\left(m_t-m_{i_3i_4j_2}\right)^2}{\sigma_t^2}\Big\},
\end{align}
where $i_1, i_2, i_3, i_4$ refer to light jets and $j_1, j_2$ refer to $b$-jets, and we take $\sigma_W=10.81$ GeV and $\sigma_t=31.01$ GeV. The ``min'' runs over all possible permutations of light jets and $b$-jets in the event. And finally, events with $\chi^2>6$ are vetoed. 

After all these cuts, the di-Higgs invariant mass distributions for both the signal and the backgrounds are shown in FIG.\,\ref{fig:bkg}. For illustration, we only show our signal with $\kappa_\lambda=1$ as represented by the black histogram, which corresponds to the SM scenario. Then by fitting these histograms, we obtain the expected confidence level scan as a function of $\kappa_\lambda$ for the $pp\rightarrow hh+jet+X$ process as shown in FIG.\,\ref{fig:cls}. There, we use $\Lambda_0$ for the significance with $\kappa_\lambda=1$ for the SM case, and $\Lambda$ that with generical $\kappa_\lambda$'s. Clearly, negative $\kappa_\lambda$'s would be excluded beyond $2\sigma$ CL by future 100\,TeV $pp$ colliders using $pp\rightarrow hh+jet+X$. Furthermore, the allowed $2\sigma$ CL range of $\kappa_\lambda$ would also shrink to $\sim[0.5, 1.7]$ compared to those in Ref.\,\cite{ATLAS:2021ifb}. Finally, the significance distributions for $pp\rightarrow hh+jet+X$ and $pp\rightarrow hh+X$ are shown in FIG.\,\ref{fig:sigdec}, where the latter is calculated using the $m_{hh}$ distributions in Ref.\,\cite{Goncalves:2018qas}. 

Additionally, we analyzed our $pp\rightarrow hh+jet+X$ events with the current experimental cuts, which replace our $p^{leading}_{T,j}>150 \rm{GeV}$ with $p^{\gamma\gamma}_T>150\rm{GeV}$ and $p^{b\bar{b}}_T>150\rm{GeV}$. And we find that about $23\%$ of the signal events which passes our cuts can not pass the current experimental cuts. And in the $250{\rm{GeV}}<m_{hh}<400{\rm{GeV}}$ region, this number is $67\%$. These numbers show clearly that the $pp\rightarrow hh+jet+X$ channel does provide extra information on $\kappa_\lambda$ that would eventually help the determination of the latter.

\begin{figure}[h]
\caption{\label{fig:measure}The $1\sigma$ (yellow) and $2\sigma$ (green) bands for $\kappa_{\lambda}$ measurement at a future $100 ~ \rm{TeV}$ $pp$ collider with $\mathcal{L}=30\,\rm ab^{-1}$. The theory predictions on the Higgs self coupling within the $1 \sigma$ uncertainty in different Higgs scenarios are also shown.}
\includegraphics[width=\linewidth]{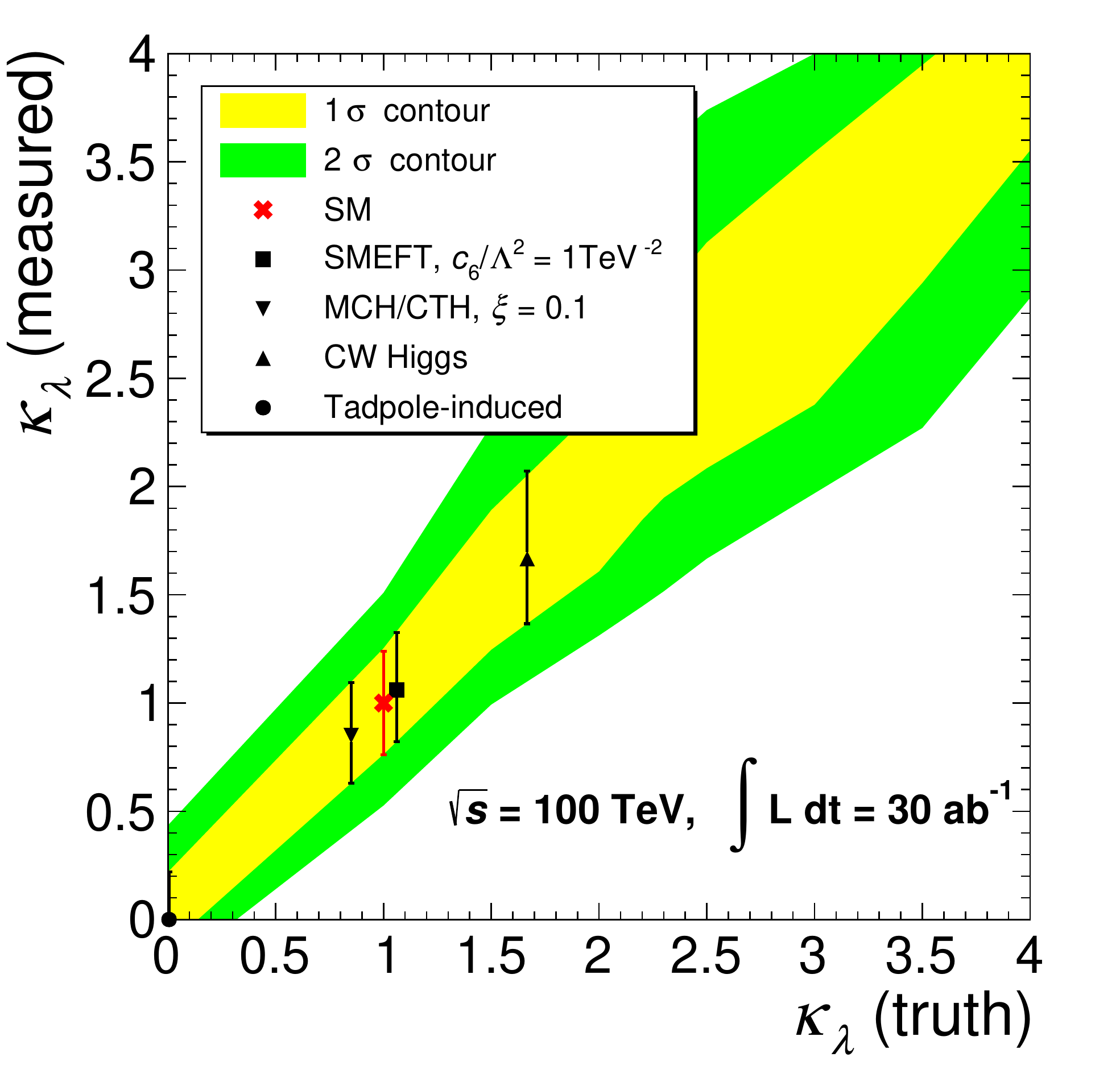}
\centering
\end{figure}

Given the sensitivity of a future 100\,TeV $pp$ collider on $\kappa_\lambda$ as just discussed, we then ask: What precision level could a future 100\,TeV $pp$ collider achieve in extracting $\kappa_\lambda$ from the data? To answer this question, we utilize our results in FIG.\,\ref{fig:sigdec} and obtain the $1\sigma$ and $2\sigma$ bands in $\kappa_\lambda$ determination at a future 100\,TeV $pp$ collider. The result is shown in FIG.\,\ref{fig:measure}, with the yellow (green) representing the $1\sigma$ ($2\sigma$) bands, respectively. Note that since negative $\kappa_\lambda$'s would be ruled out beyond $2\sigma$ CL as discussed above, we only present our result for positive $\kappa_\lambda$'s in FIG.\,\ref{fig:measure}. On the other hand, as seen from FIG.\,\ref{fig:measure}, the $1\sigma$ and $2\sigma$ bands are broader for larger $\kappa_\lambda$'s mainly due to the significance drop when $\kappa_\lambda$ increases, which is already seen in FIG.\,\ref{fig:sigdec}. This significance drop mainly seeds in the deconstructive interference between FIG.\,\ref{fig:FD}\,(a,b) and (c,d) as similarly in the $pp\to hh$ case, which in turn is guaranteed by the low-energy theorem\,\cite{Hagiwara:1989xx,Kniehl:1995tn}. Finally, as depicted in FIG.\,\ref{fig:measure}, we find the $1\sigma$ uncertainty of $\kappa_\lambda$ would be around 0.2 (1.05) in the small (large) $\kappa_\lambda$ region, mainly as a result of statistical uncertainties.

We also show the typical benchmark points for each kinds of the Higgs scenarios in FIG.\,\ref{fig:measure}: the SM, SMEFT with $c_6/\Lambda^2 = 1$ TeV$^{-2}$, the MCH/CTH with $\xi = 0.1$, the CW Higgs and the tadpole induced Higgs, in which the Higgs self couplings are taken from TAB.~\ref{tab:coupling}. We find that given the 30 ab$^{-1}$ luminosity data, it is likely to distinguish the non-decoupling scenarios (CW and Tadpole-induced) from the SM-like scenarios (SM, SMEFT, and MCH/CTH). On the other hand, it is hard to distinguish scenarios inside the SM-like scenarios, such as between the SM and the SMEFT and MCH/CTH ones. This is because the Higgs couplings to the gauge bosons and the SM fermions put tight constraints on the parameters $c_6/\Lambda^2$ and $\xi$ in such scenarios. Note that the result shown in FIG.\,\ref{fig:measure} only utilize the di-Higgs plus jet data, while combining this data and the future di-Higgs data might provide some possibility to distinguish scenarios between the SM and the SMEFT and MCH/CTH ones. 

\section{Conclusions}
\label{sec:con}
Higgs self couplings are of fundamental importance to our understanding of nature. In this letter, we propose to use the $pp\rightarrow hh+jet+X$ channel as a complementary probe of Higgs self couplings. Compared to the conventional searches with $pp\rightarrow hh+X$, we require the existence of an extra hard jet in the final state to suppress the QCD background and improve $\kappa_\lambda$ extraction in the low $m_{hh}$ region, where it is most sensitive to new physics. Due to the limited statistics at the LHC even in its high-luminosity era, we work instead at a future 100\,TeV $pp$ collider. We find that: 
\begin{itemize}
\item While the most recent analysis from ATLAS\,\cite{ATLAS:2021ifb} still permits negative $\kappa_\lambda$ at 95\% CL, it would generically be disfavored beyond $2\sigma$ CL using our signal $pp\rightarrow hh+jet+X$ at a future 100\,TeV $pp$ collider. This can be seen from our FIG.\,\ref{fig:cls};
\item Compared to the most recent results reported by ATLAS in\,\cite{ATLAS:2021ifb} using $pp\rightarrow hh$, the $2\sigma$ allowed interval of $\kappa_\lambda$ by utilizing our signal would be improved from $-1.5\le\kappa_\lambda\le6.7$ to $0.5\lesssim\kappa_\lambda\lesssim1.7$, corresponding to an improvement by almost a factor of 5;
\item A full kinematic analysis of $pp\to hh$ at the LHC with $\mathcal{L}=3\rm\,ab^{-1}$ in Ref.\,\cite{Kling:2016lay} leads to $-0.2\le\kappa_\lambda\le2.6$, and Ref.\,\cite{Goncalves:2018qas} further improved that to $0.3\le\kappa_\lambda\le1.3$ ($0.90\le\kappa_\lambda\le1.09$) for a high-energy $pp$ collider with $\sqrt{s}=27\,(100)$\,TeV. Our results would improve the result in\,\cite{Kling:2016lay}  by a factor of $\sim1.7$, and would be comparable to that in\,\cite{Goncalves:2018qas}.
\item Our result is not as good as the result shown in \cite{Taliercio:2022maa}. This is because in our analysis, we only use the di-Higgs plus one hard jet events since we focus on investigating the information carried by these signal events. These events, although carries information of the low $m_{hh}$ distribution, are only small part of the signal events. A combination with regular signal events will highly increase the total event number and suppress the statistic uncertainty. However, we show that these signal events are helpful to study the low $m_{hh}$ distribution and thus the strength of the self-interaction of the Higgs boson, and a lot of them are missed in current analysis. We suggest our experimentalists colleagues consider to add them back to their signal events. 
\end{itemize}
Finally, we present the prospect of the precision determination for $\kappa_\lambda$ at a future 100\,TeV $pp$ collider in FIG.\,\ref{fig:measure}. We find that, depending on the magnitude of $\kappa_\lambda$, its $1\sigma$ uncertainty at a future 100\,TeV $pp$ collider could be around 0.2 (1.05) for small (large) $\kappa_\lambda$'s. Given the 30 ab$^{-1}$ luminosity data, we find that it is likely to distinguish the non-decoupling scenarios (CW and Tadpole-induced) from the SM-like scenarios (SM, SMEFT, and MCH/CTH). On the other hand, it is hard to distinguish scenarios inside the SM-like scenarios, such as between the SM and the SMEFT and MCH/CTH ones. 

A few comments are in order. First, a machine-learning based approach on the same study is expected to improve our results by a few, and it would be desirable to see its impact on further distinguishing different theory scenarios. Second, we expect that in future a combined analysis among different channels would finally help determine the Higgs self couplings and unveil the nature of the Higgs boson.

\section*{Acknowledgments} 
We thank Yong Du for his valuable contribution at the early stage of this project, and the HPC Cluster of ITP-CAS for the computation support. J.-H.Y. is supported by the National Science Foundation of China under Grants No. 12022514, No. 11875003 and No. 12047503, and National Key Research and Development Program of China Grant No. 2020YFC2201501, No. 2021YFA0718304, and CAS Project for Young Scientists in Basic Research YSBR-006, the Key Research Program of the CAS Grant No. XDPB15. K. C. and H. Z. are supported by the National Science Foundation of China under Grants No. 12075257 and No. 12235001, and the funding from the Institute of High Energy Physics, Chinese Academy of Sciences with Contract No. Y6515580U1, and the funding from Chinese Academy of Sciences with Contract No. Y8291120K2. J.-H.Y. and H. Z. are pleased to recognize the support and the hospitality of the Center for High Energy Physics at Peking University.

\bibliographystyle{utcaps_mod}
\bibliography{main}
\end{document}